\documentclass[aps,pra,twocolumn,showpacs,floatfix]{revtex4}
\usepackage{graphicx}
\usepackage{amsmath}
\usepackage[usenames,dvipsnames]{color}
\usepackage[percent]{overpic} 

\def\mathbi#1{\textbf{\em #1}}

\begin{document}

\title{Magnetic Feshbach resonances in collisions of non-magnetic closed-shell $^1\Sigma$ molecules}
\author{Alisdair O. G. Wallis}
\affiliation{Department of Chemistry, University of British Columbia,
Vancouver, British Columbia, V6T~1Z1, Canada}
\author{Roman V. Krems}
\affiliation{Department of Chemistry, University of British Columbia,
Vancouver, British Columbia, V6T~1Z1, Canada}

\date{\today}

\begin{abstract}
Magnetic Feshbach resonances play a central role in experimental research of atomic gases at ultracold temperatures, as they allow one to control the microscopic interactions between ultracold atoms by tuning an applied magnetic field. These resonances arise due to strong hyperfine interactions between the unpaired electron and the nuclear magnetic moment of the alkali metal atoms.  A major thrust of current research is to create an ultracold gas of diatomic alkali-metal molecules in the ground rovibrational state of the ground electronic $^1\Sigma$ state. Unlike alkali metal atoms, $^1\Sigma$ diatomic molecules have no unpaired electrons.
However, the hyperfine interactions of molecules may give rise to magnetic Feshbach resonances. We use quantum scattering calculations to study the possible width of these resonances. 
%Due to the computational difficulty of the problem, our calculations employ significantly restricted basis sets and model interaction potentials so the results cannot be used as quantitative predictions for a specific molecular system.  However,
Our results show that the widths of magnetic Feshbach resonances in ultracold molecule-molecule collisions for $^1\Sigma$ molecules may exceed 1 milliGauss, rendering such resonances experimentally detectable. We hope that this work will stimulate the experimental search of these resonances.

%We present quantum scattering calculations illustrating that nuclear spin relaxation and Feshbach resonances in ultracold molecule-molecule collisions for $^1\Sigma$ molecules in a magnetic field are probable. These resonances are mediated by an interplay of the couplings between different rotational states due to the anisotropy of the intermolecular interaction potential and the interactions of the nuclear spins with the rotational motion of the molecules. 

%We show that certain Zeeman states of $^1\Sigma$ molecules exhibit broad resonances and weak inelastic scattering, which makes these states ideal for experiments exploiting magnetic field control of ultracold molecular gases. 
\end{abstract}


\maketitle

%\fbox{\parbox{0.95\linewidth}{{\em Note for copy editor:} We have
%been very careful to make correct use of Roman and italic
%subscripts and superscripts, with Roman for abbreviations and
%italic for mathematical indices. Please do not change all our
%subscripts and superscripts to italic.}}

%{\it 
%From the referees report, some of the key issues we need to address are
%\begin{itemize}
%\item why are Feshbach resonances between closed-shell $^1\Sigma$ molecules surprising
%\item ``The paper is not very well written, and the figures are poorly explained."

%\item elucidate the mechanism for the Feshbach resonances

%\item details of the potential energy surface, why we use NH-NH

%\item more emphasis on what our results can predict and why they are interesting (``interesting problem using drastic approximations")
%\begin{itemize}
%\item the existence of resonances
%\item lower limit
%\item mechanism
%\end{itemize}

%\end{itemize}
%}

\vspace{1.5 cm}

%
% Figures
%
% ----
% figure 1, 87Rb133Cs Hyperfine structure
% ----
% figure 2, spin relaxation cross sections (energy dependence)
% ----
% figure 3, spin-relaxation cross sections with various hyperfine parameters turned off
% ----
% figure 4/ table 1, resonance widths 
%

% ----

%
%  Introduction
%
% 1) Why are Feshbach resonances a surprise in 1\Sigma molecules (no electron spin)
%		- what is the mechanism for alkali metal atoms
%		- JMH work
%		- what do we do for the first time/what is surprising about our results to justify a PRL?
%		  the first few paragraphs need to answers these questions
%
% Interest in creating 1sigma ground state molecules KRb, RbCs (stable), 
% magnetic feshbach resonances are used to control collisional properties
% \cite{ Rev. Mod. Phys. 82, 1225�1286 (2010) Feshbach resonances in ultracold gases}
% but can magnet FB resonances be used to control 1sigma molecular collisions
% mechanism for FBres in alkali-metal atoms is..., recently JMH RbSr, LiYb FBres
% but what about 1sigma+1sigma - no electronic spin
%
% The purpose of this letter is present the first calculations of hyperfine relaxation in
% molecule-molecule collisions and explore the possibility of tuning magnetic FBres
% in RbCs(1sigma) molecules
%
% RbCs molecules are subject of much current research

\section{Introduction}

Much of the success of experimental research with ultracold atomic gases is due to the possibility of tuning the scattering length of ultracold alkali metal atoms by means of magnetic Feshbach resonances \cite{Chin:feshbach:2010}. The scattering length near a Feshbach resonance undergoes rapid variation as a function of the magnetic field, which has been used for experimental studies of Bose--Einstein condensation (BEC) \cite{Inouye:1998,Timmermans:1999}, bosonic superfluidity \cite{Sengupta:2005}, quantum magnetism \cite{Duan:2003}, many-body spin dynamics \cite{Widera:2008}, Efimov states \cite{Kraemer:2006}, Bardeen--Cooper--Schrieffer (BCS) superfluidity \cite{Kinast:2004} and the BEC--BCS crossover \cite{Regal:res-cond:2004,Bourdel:2004}. A major thrust of current experiments is to extend this work to ultracold molecules \cite{Krems:book:2009}. Of particular interest are polar alkali metal dimers in the ground $^1\Sigma$ electronic state produced by %photoassociation of ultracold atoms. % \cite{Fioretti:1998}. 
 magnetoassociation of ultracold atoms into a weakly bound state followed by the coherent transfer into the absolute ground state via stimulated Raman adiabatic passage (STIRAP).
Several experiments have recently demonstrated the creation of ultracold KRb molecules \cite{Ni:KRb:2008} in the  
ground electronic and ro-vibrational state
and rapid progress is being made with RbCs \cite{Cho:towardsRbCs:2011,Takekoshi:RbCs:2012}
%, as well as with 
and
other alkali metal 
%diatomic molecules 
dimers
\cite{Deiglmayr:LiCs:2011,Zabawa:NaCs:2011,Wang:NaRb:2013,Ferber:KCs:2013}.

Polar molecules offer long-range dipolar interactions, which can be used for new exciting applications such as the study of dipolar crystals \cite{Pupillo:book:2009} and ultracold controlled chemistry \cite{Krems:PCCP:2008, ospelkaus:krb:2010}. While the long-range interactions between polar molecules can be controlled by an external electric field \cite{Ni:dipolar:2010,Quemener:2010}, many applications, such as quantum simulation of spin-lattice models \cite{Micheli:2006} or the study of dipolar effects on BEC and the BEC - BCS crossover \cite{Baranov:2002}, require independent control over short-range and long-range interactions. The short-range interactions could, in principle, be controlled by magnetic Feshbach resonances. 

In order for atoms or molecules to exhibit magnetic Feshbach resonances, it is necessary that (i) the collision partners exhibit the Zeeman effect; and (ii) the different Zeeman states be coupled by the interaction potential of the colliding partners. The Zeeman effect of alkali metal atoms is determined by both the electron and nuclear spins. Due to strong hyperfine interactions between the unpaired electron and the nuclear magnetic moment in these atoms, each Zeeman state is a linear combination of the electron and nuclear spin states. If expressed in the basis of the Zeeman states, the matrix of the interaction potential between two alkali metal atoms contains off-diagonal matrix elements proportional to the energy difference between the singlet ($^1\Sigma$) and triplet ($^3\Sigma$) electronic states  of the collision complex \cite{Li-Krems:2007}. These matrix elements give rise to magnetic Feshbach resonances.

Consider now closed-shell atoms with non-zero nuclear spin - such atoms exhibit the Zeeman effect due to the non-zero magnetic moment of the nucleus. However, the interaction potential between the colliding atoms does not perturb the Zeeman structure. Because the Zeeman states do not depend on the electronic degrees of freedom, 
the matrix of the interatomic interaction potential in the basis of the nuclear spin states is diagonal and the collision properties of atoms in different Zeeman states are identical. Such atoms  do not exhibit magnetic Feshbach resonances.

The ultimate goal of the magneto- and photo- assoication experiments is to produce molecules in the absolute ground state. If chemically non-reactive \cite{Zuchowski:rxns:2010}, molecules in the absolute ground state are collisionally stable. The lowest-energy state of $^1\Sigma$ diatomic molecules is characterized by zero electronic spin angular momentum and zero rotational angular momentum of the nuclei. If excited rotational states are ignored, $^1\Sigma$ molecules in the ground rotational state can be viewed as closed-shell atoms. When placed in a magnetic field, $^1\Sigma$ molecules exhibit the Zeeman effect due to non-zero magnetic moments of the nuclei. However, the nuclear spins are uncoupled from the electronic degrees of freedom. To first order, the nuclear spins are also uncoupled from the rotational or translational motions of  the molecules.
% ---
% original
%To first order, the nuclear spins are also uncoupled from the rotational or translational motions of  the molecules and it is currently believed that the Zeeman shifts in $^1\Sigma$ molecules cannot be exploited for tuning the collision properties of molecules in the ground rotational state. 
% ---
% new
%{\color{blue}
%To first order, the nuclear spins are also uncoupled from the rotational or translational motions of  the molecules. Currently there have been no attempts to examine the explicit couplings that could lead to magnetic Feshbach resonances or calculate the widths of the possible magnetic Feshbach resonances from coupled channel scattering calculations. 
%}

The presence of closely-spaced rotational states makes $^1\Sigma$ molecules different from closed-shell atoms.  The electronic interaction between molecules couples different rotational states of the molecules. Because the Zeeman structure of molecules in different rotational states is different, the interplay of the couplings induced by the molecule - molecule interaction potential and the interactions induced by molecular rotations may give rise to magnetic Feshbach resonances that can be used to control ultracold collisions of $^1\Sigma$ molecules. The experimental feasibility of magnetic field control of molecule - molecule collisions depends on the widths of these resonances. 
  While the above discussion suggests that these resonances must be narrower than the magnetic resonances in collisions of open-shell species, there have been no attempts to calculate the widths of magnetic resonances in molecule - molecule collisions.

In this work, we present quantum scattering calculations 
illustrating that Feshbach resonances in ultracold 
molecule-molecule collisions for $^1\Sigma$ molecules in a magnetic field can have experimentally detectable widths. To elucidate the mechanism of the Feshbach resonances, we explore the role of the different hyperfine interactions coupling different Zeeman states. The calculations are performed for collisions of molecules  having the same structure of hyperfine and rotational levels as RbCs in the ground $^1\Sigma$ electronic state.  Since the results of scattering calculations for molecular collisions at ultracold temperatures are very sensitive to the interaction potential, we perform calculations with a series of model potential surfaces.  The inclusion of the nuclear spin states in quantum scattering calculations greatly increases the computational difficulty of the problem, forcing us to restrict the basis for the scattering calculations to three or four rotational states for each molecule.
These limitations do not allow us to predict the widths of Feshbach resonances  for any specific molecule. Instead, the main question we aim to answer is,  can the hyperfine interactions similar to those in RbCs($^1\Sigma$) -- in principle -- lead to experimentally detectable widths of magnetic Feshbach resonances?

\section{Theory}
% -----
% figure 1, 87Rb133Cs Hyperfine structure
%
\begin{figure}
\includegraphics[width=1.0\linewidth]{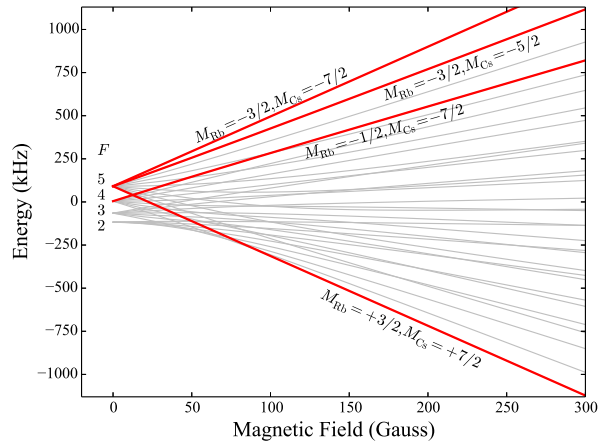}
\caption{(color online) The hyperfine structure of the $^{87}$Rb$^{133}$Cs(X$^1\Sigma^+$) rotational ground-state. The hyperfine states considered here are highlighted in red.
\label{fig:RbCs_hf}
}
\end{figure}
% -----

In order to calculate the Zeeman energy levels of the $^{87}$Rb$^{133}$Cs(X$^1\Sigma^+$) molecule, we diagonalize the following Hamiltonian \cite{Brown:2003}:
\begin{equation}\label{eqn:Ham}
\hat{H}=B_{\rm rot}{\mathbi{N}}^2 + \hat{H}_\text{hf} + \hat{H}_\text{Zeeman},
\end{equation}
where $B_{\rm rot}$ is the rotational constant of the molecule, ${\mathbi{N}}$ is the molecular angular momentum excluding nuclear spins, the operator $\hat{H}_\text{hf}$ describes the hyperfine interaction \cite{Aldegunde:polar:2008} and $\hat{H}_\text{Zeeman}$ describes the response of the molecule to an external magnetic field. 

 Following Aldegunde \emph{et al.} \cite{Aldegunde:polar:2008}, we write the hyperfine interaction as
\begin{equation}
\hat{H}_\text{hf}=
\sum_{i=1}^2  \mathbi{V}_i\cdot\mathbi{Q}_i 
+\sum_{i=1}^2c_i \mathbi{N}\cdot\mathbi{I}_i 
+c_3\mathbi{I}_1\cdot \mathbi{T}\cdot\mathbi{I}_2
+c_4\mathbi{I}_1\cdot\mathbi{I}_2,
\label{hf}
\end{equation}
where $\mathbi{I}_i$ is the nuclear spin of nucleus $i$. 
The first term in Eq. (\ref{hf}) is the electric-quadrupole interaction, which represents the interaction of the nuclear quadrupole moment $\mathbi{Q}_i$ with the electric field gradient $\mathbi{V}_i$ at nucleus $i$. The second term is the nuclear spin-rotation interaction between the magnetic moment of the nuclear spin and the magnetic moment created by the molecular rotation. The third and fourth terms describe the tensor and scalar parts of the nuclear spin-spin interaction respectively, the tensor $\mathbi{T}$ describes the direct dipolar interaction and the anisotropic part of the indirect interaction, which occurs via the electronic distribution \cite{Bryce:2003,Brown:2003}. 
%The constants in the hyperfine Hamiltonian have been calculated by Aldegunde \emph{et al.} \cite{Aldegunde:polar:2008} and 
The constants in the hyperfine Hamiltonian are borrowed from Ref. \cite{Aldegunde:polar:2008}. They have the following values: $c_{^{87}\rm Rb}=98.4$ Hz, $c_{^{133}\rm Cs}=194.1$ Hz, $c_3 = 192.4$ Hz and $c_4 = 17345.4$ Hz, the nuclear quadrupole moments are: $(eQq)_{^{87}\rm Rb} = -0.872$ MHz and $(eQq)_{^{133}\rm Cs} = 0.051$ MHz.

The interaction with the magnetic field is a sum of the rotational and nuclear contributions
\begin{equation}
\hat{H}_\text{Zeeman} = 
-g_r\mu_N\mathbi{N}\cdot\mathbi{B} -
\sum_{i=1}^2g_i\mu_N\mathbi{I}_i\cdot\mathbi{B}(1-\sigma_i),
\label{zeeman}
\end{equation}
 where $g_r$ and $g_i$ are the rotational and nuclear $g$-factors and $\sigma_i$ is a nuclear screening factor. The calculated values  \cite{Aldegunde:polar:2008} of the constants are $g_r=0.0062$, $g_{^{87}\rm Rb}=1.834$, $g_{^{133}Cs}=0.738$, $\sigma_{^{87}\rm Rb}=3531$ ppm and $\sigma_{^{133}\rm Cs}= 6367$ ppm.

Following Aldegunde \emph{et al.} \cite{Aldegunde:polar:2008}, we diagonalize the total Hamiltonian using the fully uncoupled space-fixed basis, $|I_\text{Rb} M_{\text{Rb}} \rangle| I_\text{Cs} M_{\text{Cs}}\rangle | N M_N\rangle$,  where $I_{^{87}\text{Rb}}=\tfrac{3}{2}$, $I_{^{133}\text{Cs}}=\tfrac{7}{2}$ and the projections of the corresponding angular momenta $M_\text{Rb}$, $M_\text{Cs}$, and $M_{N}$ are defined with respect to the direction of the magnetic field vector. The hyperfine states of $^{87}$Rb$^{133}$Cs(X$^1\Sigma^+$) in the ground rotational state are shown in Figure~\ref{fig:RbCs_hf}. 
At zero magnetic field, the energy levels are labeled by the quantum number of the total angular momentum $\mathbi{F}=\mathbi{I}_{\rm Rb}+\mathbi{I}_{\rm Cs}+\mathbi{N}$. At high magnetic fields, the energy levels can be labeled by the nearly good quantum numbers $M_{\rm Rb}$ and $M_{\rm Cs}$. Recent experiments demonstrated that ultracold alkali metal dimers can be selectively prepared in any Zeeman state of the ground rotational state \cite{Ospelkaus:2010}. In the present work, we examine the strength of collision-induced couplings between the Zeeman states shown in Figure~\ref{fig:RbCs_hf}. These couplings lead to nuclear spin relaxation and magnetic Feshbach resonances in molecule - molecule collisions.

To evaluate the outcome of collisions, we follow the approach described in Refs. \cite{Krems:mfield:2004,Tscherbul:molmol:2009}.  The total scattering wavefunction, for molecules $A$ and $B$, is expanded in the uncoupled space-fixed basis $|\tau_A\tau_BLM_L\rangle$, where $|\tau_i\rangle= | I_\text{Rb}^{(i)} M_{I_\text{Rb}}^{(i)} I_\text{Cs}^{(i)} M_{I_\text{Cs}}^{(i)} N^{(i)} M_N^{(i)} \rangle$, for molecule $i$, and $L$ is the end-over-end rotational angular momentum of the molecule - molecule collision complex. This basis is then symmetrized with respect to the interchange of two identical molecules \cite{Green:1975,Alexander:symmetry:1977}. The total angular momentum is not conserved for collisions in a magnetic field. However, the total angular momentum projection $M_\text{tot}$ and the parity are conserved and the Hamiltonian can be block diagonalized. For each value of $M_\text{tot}$ and parity, we reduce the Schr\"{o}dinger equation to a set of coupled-channel equations. The solutions of these equations are propagated outwards over a radial grid of molecule - molecule separation using the improved log-derivative method of Manolopoulos \cite{Manolopoulos:1986}.  We divide the radial grid into short ($4$ to $80$ a$_0$) and long (80 to 1000 a$_0$) range regions and use the grid step sizes of 0.0305 a$_0$ (for the short range region) and 1 a$_0$ (for the long range region). At 1000 a$_0$, the log-derivative matrix is transformed into the basis of asymptotic eigenvectors \cite{Krems:mfield:2004} and matched to the appropriate scattering boundary conditions \cite{Johnson:1973} to obtain the  $S$-matrix. The scattering cross sections are then calculated from the $S$-matrix elements. 

The integration time of the scattering differential equations scales as the third power of the number of channels (number of coupled equations). Incorporating the hyperfine structure into molecule-molecule scattering calculations creates hundreds to thousands of additional channels making the calculations unfeasible. Approximations are thus required to make calculations feasible and allow the investigation of hyperfine structure effects in molecular collisions.

\section{Summary of approximations}

 The accuracy of the coupled-channel calculations depends on the number of basis states and on the accuracy of the molecule - molecule interaction potential employed for the computations. In this section we discuss the approximations used to make our calculations feasible, the justification for these approximations and the expected effect of the simplifications on the computed observables.

% i) fully converged calc's not possible ii) therefore to investigate the system approximations are required

%

The highest order interaction, that is predominantly responsible for Feshbach resonances, is 
a combination of the second order couplings due to the nuclear quadrupole interaction and the molecule - molecule interaction potential. To first approximation, we propose to neglect all states that are not coupled to a desired initial channel via these second order couplings.
Practically, this requires neglecting channels that differ from a specified initial channel by $|\Delta M_{I_\text{tot}}| > 2 N_\text{max}$, where $M_{I_\text{tot}} = M_{\text{Rb}}^{(A)}+M_{\text{Cs}}^{(A)}+M_{\text{Rb}}^{(B)}+M_{\text{Cs}}^{(B)}$ and $N_\text{max}$ is the largest value of the rotational angular momentum included in the calculation.
Neglecting these states approximately halves the number of channels in the calculation. 
For example, this approximation reduces the number of coupled equations encompassing the molecular states with $N\le3$ and $L\le2$ for $M_\text{tot}=10$ from 4238 to 2060. 
We have confirmed  that this approximation affects the calculated cross sections and resonance widths by less than 1 $\%$.

 Even with this reduction of the basis set, it is impossible to converge the cross sections for molecule - molecule scattering with respect to the number or rotational states in the basis. Most of the calculations reported here were performed with the basis set restricted to three rotational states of each molecule ($N \le 2$) and three angular momenta  ($L \le 2$) for the end-over-end rotation of the collision complex. This basis set allows us to include all first-order and second-order couplings due to the hyperfine interactions. Such restricted basis sets were previously used to draw qualitative conclusions about the dynamics of electronic spin relaxation in molecule - molecule collisions \cite{Janssen:NHmagnetic:2011,Janssen:fieldfree:2011,Janssen:NHdipolar:2011}.

The following section reports the cross sections for nuclear spin relaxation and the widths of magnetic Feshbach resonances. It is necessary to consider the effect of the rotational basis set truncation on these computed observables.  Including more rotational states in the basis will lead to more (indirect) couplings between the initial nuclear spin state and other Zeeman states. Therefore, it should be expected that the calculations with larger basis sets will -- in general -- yield larger cross sections for nuclear spin relaxation and magnetic Feshbach resonances with larger widths.  
%In specific cases, the inclusion of more basis states may lead to interference effects, decreasing the resonance widths. However, our goal is not to consider specific resonances. Rather, 
Our aim is to answer the general question of whether the hyperfine interactions are strong enough to lead to observable inelastic cross sections and Feshbach resonances. We argue that if the hyperfine interactions are strong enough in a small basis allowing for first- and second-order couplings, they must be equally or more effective when higher-order interactions are included.

The third simplication made in the present work concerns the potential energy surface. Although the interaction energy of the RbCs - RbCs complex has been calculated for selected geometries~\cite{Tscherbul:beams:2008,Byrd:2012},
there is currently no full potential energy surface available for two alkali metal dimers in the ground electronic state. As a starting point for our calculations, we use the quintet potential surface recently computed for NH($^3\Sigma^-$)-NH($^3\Sigma^-$) collisions by Janssen \emph{et al.} \cite{Janssen:NHPES:2009}. The interaction potential between polar alkali metal dimers is expected to be more deeply bound and anisotropic than the NH-NH surface \cite{Tscherbul:beams:2008,Byrd:2012}. 
 For example, the interaction energy of the RbCs - RbCs complex is approximately 1000 cm$^{-1}$ \cite{Tscherbul:beams:2008} at the minimum of the potential energy surface, whereas that of the NH - NH complex is 675 cm$^{-1}$ \cite{Janssen:NHPES:2009}. 
Larger interaction strength and stronger anisotropy must lead to stronger couplings.  Therefore, the cross sections for the nuclear spin relaxation and Feshbach resonance widths computed with the NH - NH surface should generally be smaller than those computed with more anisotropic surfaces.

In summary, the combined effect of the basis set truncation and the use of the NH - NH surface leads to fewer and less strong off-diagonal couplings. Therefore,  our calculations of the nuclear relaxation cross sections and the resonance widths should be considered as lower bounds of general predictions. In addition, we present results for a range of scaled interaction potentials. 
 Quantum calculations of scattering observables for ultralow collision energies are usually sensitive to small variations of the interaction potential. As recommended in Refs. 
\cite{Janssen:fieldfree:2011,Wallis:LiNH:2011,Cui:2013}, the results must be calculated and presented for a range of potential surfaces. Averaging over multiple potential surfaces reduces the basis set trunction error \cite{Janssen:fieldfree:2011,Wallis:LiNH:2011,Cui:2013} and yields an expected range of the calculated observables.
 We multiply the NH - NH potential by a constant factor in order to scale the interaction strength over a range of values up to the predicted depth of the RbCs-RbCs potential~\cite{Tscherbul:beams:2008,Byrd:2012}. The long range region of the surface most relevant for ultracold collisions is determined by the electric dipole moment of the interacting molecules. The difference between the dipole moments of NH (1.39 D \cite{Scarl:NHdipole:1974}) and RbCs (1.25 D \cite{Aldegunde:polar:2008}) is well within the range of the scaling.

%{\color{red} how do we classify the anisotropy of a 4 dimensional $(R,\theta_A,\theta_B,\phi)$ surface, and obtain the ratio of the rotational constant to the anisotropy, $B_{\rm rot}$(RbCs) = 0.511 GHz.}

\section{Calculation Results}

In the absence of the perturbatively small hyperfine interaction Eq. (\ref{hf}), the states of a $^1\Sigma$ molecule are characterized by the quantum number $N$ of the rotational angular momentum.  
If two $^1\Sigma$ molecules were initially prepared and remained during a collision in the rotational ground state $N=0$, the nuclear spin states would be completely decoupled from the translational motion of the molecules. The molecules would behave as closed-shell atoms so no Feshbach resonances or collision-induced transitions between different Zeeman states would be possible. 
However, as can be seen from Eq. (2), the hyperfine interaction includes tensors acting on the rotational states with non-zero $N$ so the hyperfine structure depends on the rotational angular momentum $N$. In states with $N > 0$, the nuclear spins are coupled to the rotational motion of the molecules. 
 The molecule - molecule interaction potential induces couplings between rotational states that modify the hyperfine interactions during collisions. These couplings may give rise to magnetic Feshbach resonances and induce the nuclear spin relaxation in molecule - molecule collisions. 

%
% SPIN RELAXATION
%
% ----
% figure 2, spin relaxation cross sections (energy dependence)
%
\begin{figure}
\includegraphics[width=0.4\textwidth]{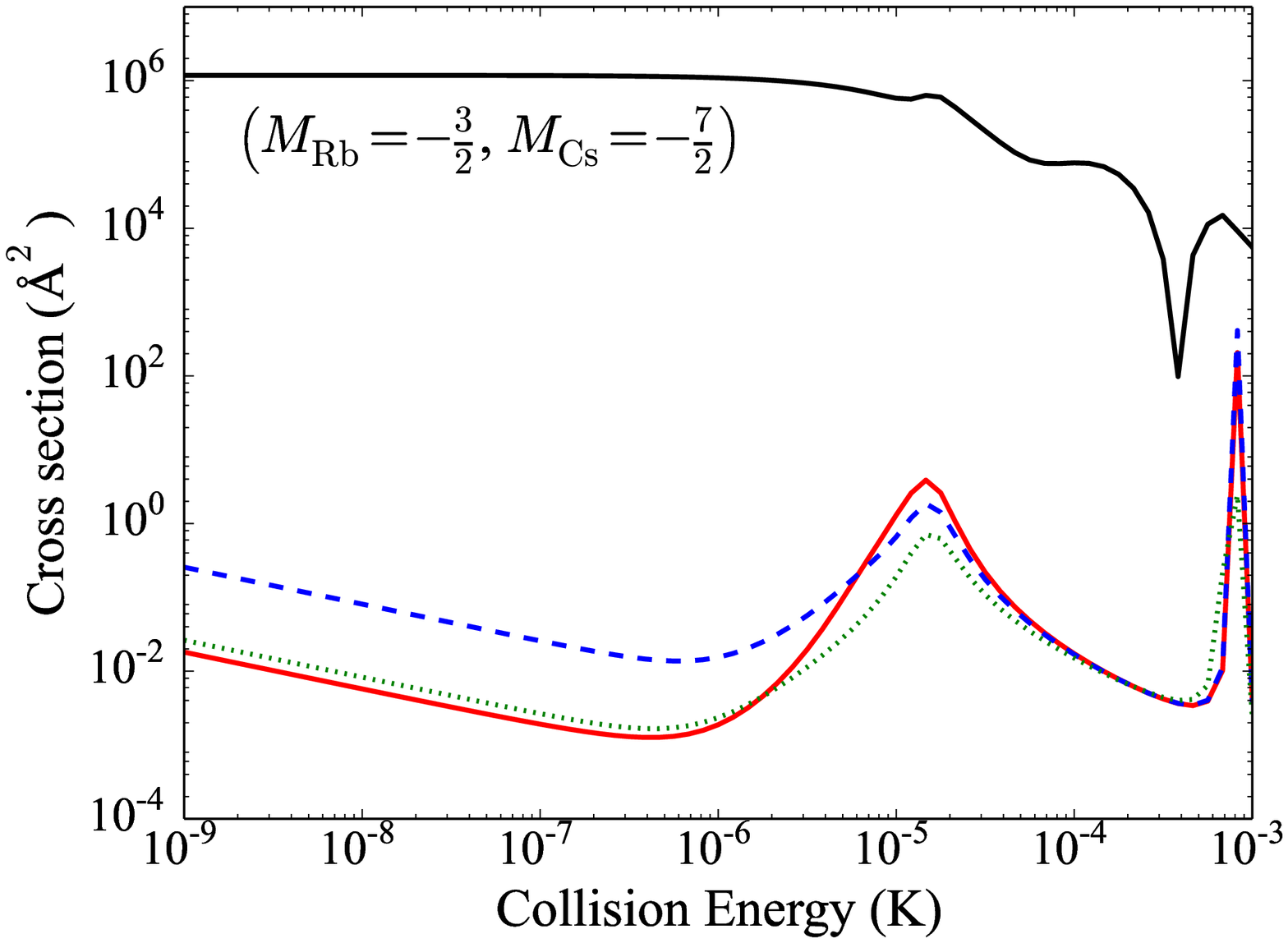}
\includegraphics[width=0.4\textwidth]{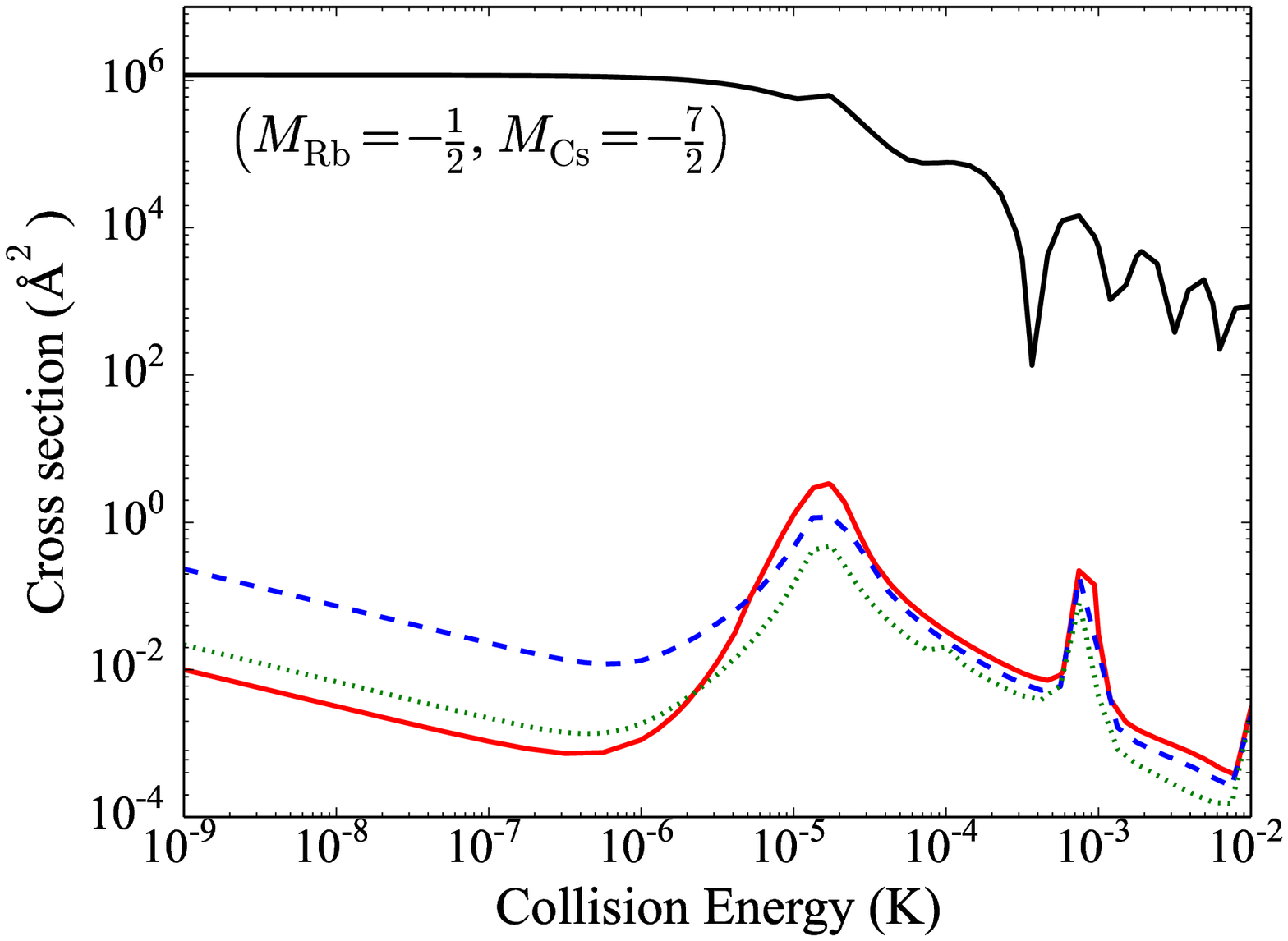}
\includegraphics[width=0.4\textwidth]{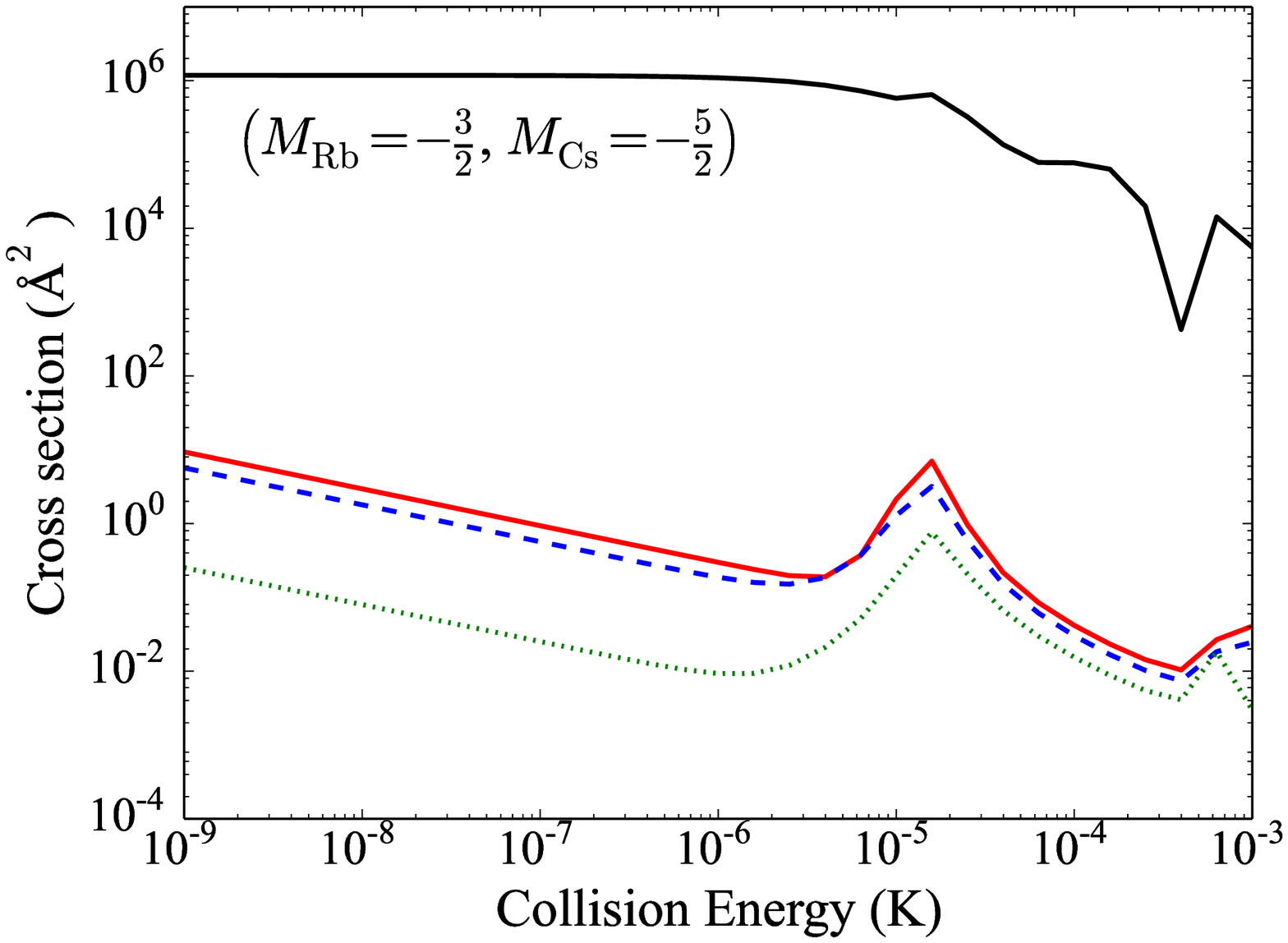}
\caption{(color online) Elastic (black-solid curves) and inelastic cross sections for two molecules in the initial states; $(M_\text{Rb} = -\tfrac{3}{2},M_\text{Cs} = -\tfrac{7}{2})$ (upper panel), $(M_\text{Rb} = -\tfrac{1}{2},M_\text{Cs} = -\tfrac{7}{2})$ (center panel), and  $(M_\text{Rb} = -\tfrac{3}{2},M_\text{Cs} = -\tfrac{5}{2})$ (lower panel) as indicated in Figure~\ref{fig:RbCs_hf}. 
The inelastic cross sections are shown at magnetic fields of 10 (red-solid curves), 100 (blue-dashed curves), 1000 Gauss (green-dotted curves).
\label{fig:spinrelax}
}
\end{figure}
%
% ----

We begin by analyzing the probability of nuclear spin relaxation of two molecules prepared in the same hyperfine state. Figure~\ref{fig:spinrelax} shows the cross sections for elastic scattering and inelastic spin-relaxation in collisions of $^{87}$Rb$^{133}$Cs molecules prepared in a number of different initial states, at a collision energy of 1 nK and at magnetic fields of 10, 100 and 1000 Gauss. The inelastic cross sections are summed over all accessible quantum states. 
The cross sections were calculated with three rotational states $N\le2$ and three partial waves  $L\le2$ in the basis set. The elastic cross section does not change appreciably with magnetic field. Therefore, we only show the elastic cross section for collisions in a magnetic field of 10 Gauss. 
% spin-stretched state, 

\begin{figure}
\includegraphics[width=0.4\textwidth]{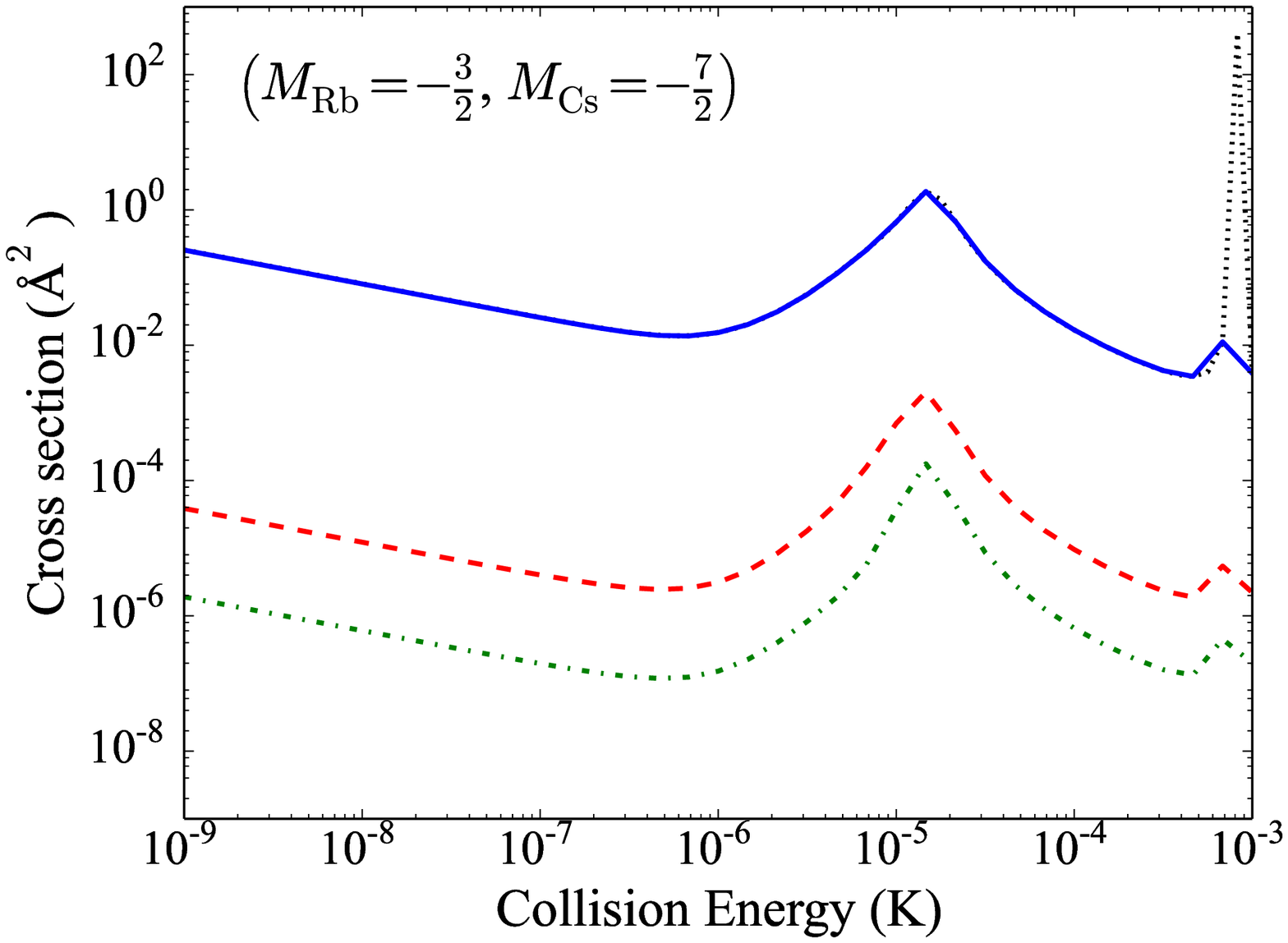}
\includegraphics[width=0.4\textwidth]{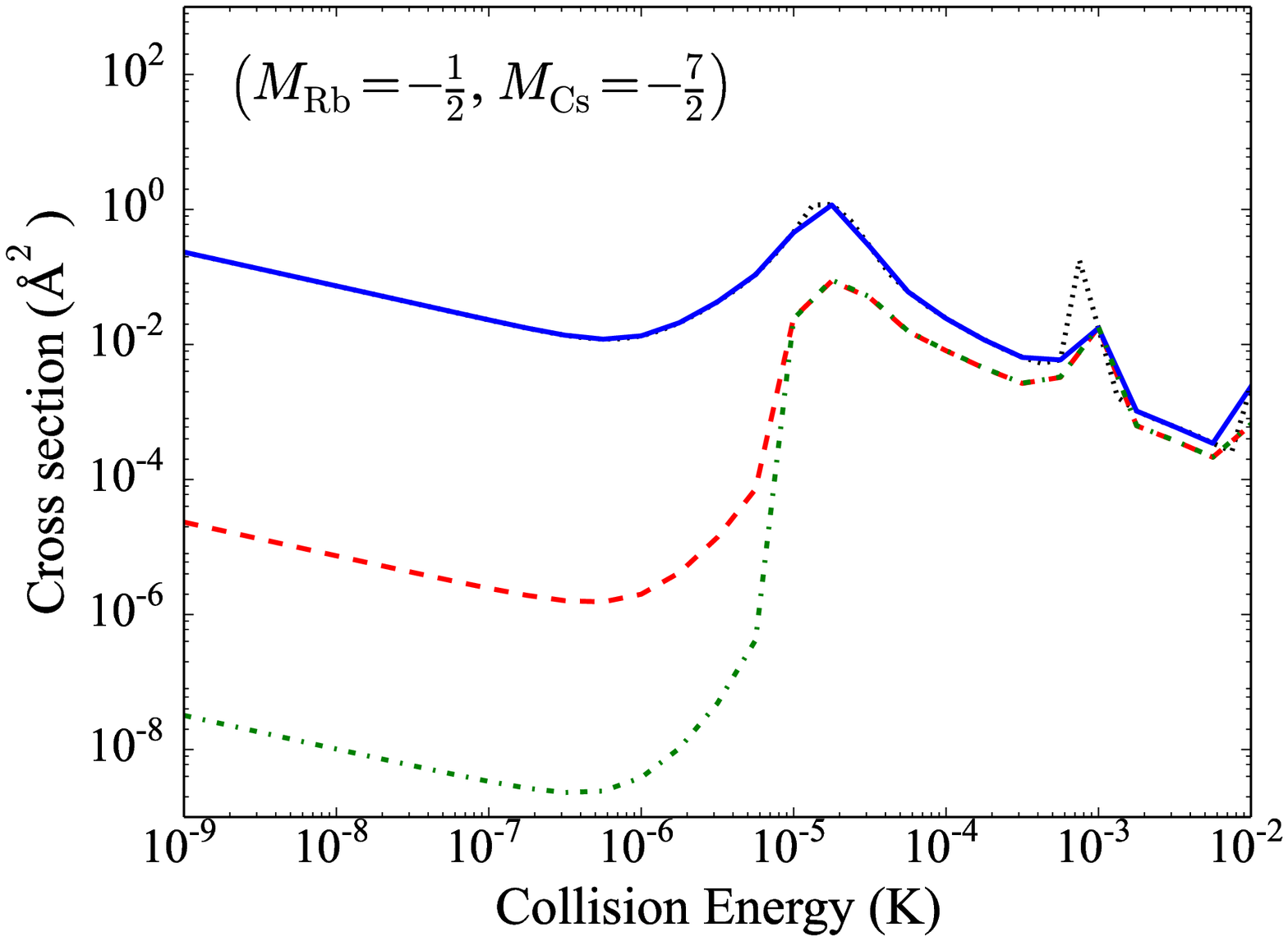}
\includegraphics[width=0.4\textwidth]{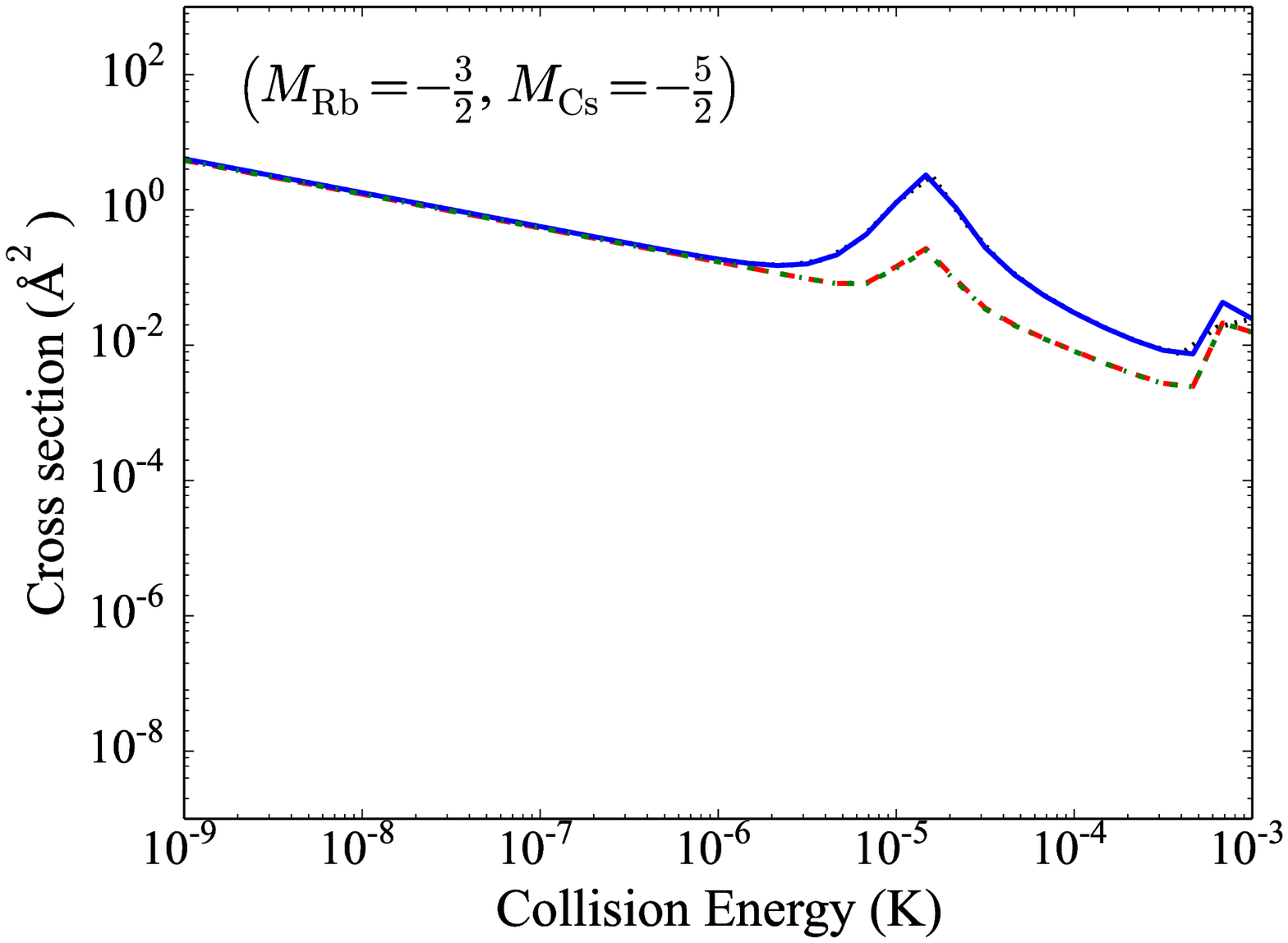}
\caption{(color online) The effect of the nuclear quadrupole moments on the inelastic cross section for two molecules in the initial state $(M_\text{Rb} = -\tfrac{3}{2},M_\text{Cs} = -\tfrac{7}{2})$ (upper panel), the initial state $(M_\text{Rb} = -\tfrac{1}{2},M_\text{Cs} = -\tfrac{7}{2})$ (middle panel), and the initial state $(M_\text{Rb} = -\tfrac{3}{2},M_\text{Cs} = -\tfrac{5}{2})$ (lower panel). 
Black-dotted curve - the full calculation, red-dashed curve - the calculation with $eQq_\text{Rb}$=0, blue-solid curve - the calculation with  $eQq_\text{Cs}$=0, green-dot dashed curve - the calculation with $eQq_\text{Rb}=eQq_\text{Cs}$=0.
%The original inelastic cross section (black-dotted) is shown along side the inelastic cross sections when $eQq_\text{Rb}$=0 (red-dashed), $eQq_\text{Cs}$=0 (blue-solid), and $eQq_\text{Rb}=eQq_\text{Cs}$=0 (green-dot dashed). 
The calculations are performed at a magnetic field of 100 Gauss.
\label{fig:mechanism}
}
\end{figure}

For molecules prepared in the maximally spin-stretched state $(M_{\rm Rb}= -3/2, M_{\rm Cs}=-7/2$, see Figure 1)  of the ground rotational state $(N^{(A)}=N^{(B)}=0)$ and in the $s$-wave collision chanel $(L=0)$ pertinent for ultracold collisions, 
  any change in the nuclear spin quantum numbers must be accompanied by a change in $L$. This leads to centrifugal barriers in outgoing collision channels that suppress inelastic relaxation \cite{Volpi:2002,Krems:mfield:2004}. 
In this case, the dominant off-diagonal coupling responsible for the nuclear spin-relaxation is induced by the  second order interplay of the nuclear quadrupole interaction and the intermolecular interaction potential. The nuclear quadrupole interaction mixes different rotational states $N$ with different $M_{i}$, where $M_{i}$ is either $M_\text{Rb}$ or $M_\text{Cs}$, and the interaction potential mixes different rotational states with the same $M_i$ but different $L$. The effect of this coupling is illustrated in Figure~\ref{fig:mechanism}, displaying the spin-relaxation cross section computed with modified Hamiltonians excluding the nuclear quadrupole moments of Rb and Cs. When the nuclear quadrupole moment of Rb is set to zero, the nuclear-spin relaxation cross section is reduced by four orders of magnitude. When the quadrupole moments of both nuclei are set to zero, the cross section is reduced by about five orders of magnitude. 
The nuclear quadrupole moment of Rb is an order of magnitude larger than that of Cs \cite{Aldegunde:polar:2008}. Therefore, setting the quadrupole moment of Cs to zero has an insignificant effect. Although the effect of the Cs quadrupole moment is small we include both the Rb and Cs quadrupole moments in the calculations of resonanace widths.

%
% non-stretched state
We have also investigated nuclear spin relaxation in collisions of molecules in other than spin-stretched states. The results presented here are for $M_\text{tot}=-8$. When (one or both) molecules are prepared in states with $|M_i| < I_i$,  there are multiple $s$-wave channels corresponding to $N^{(A)}=N^{(B)}=0$. Interestingly, when both molecules are in the state with $M_{\rm Rb}= -1/2$ and  $M_{\rm Cs}=-7/2$ (see Figure 1), there are no energetically accessible $s$-wave inelastic channels in the limit of zero collision energy (e.g., at 1 nK) and the cross section for the nuclear spin-relaxation is similar to that for collisions of molecules in the state with $M_i = I_i$ for both nuclei. This can be seen in Figure~\ref{fig:mechanism}. However, at around 10 $\mu$K, new $s$-wave channels become open and the nuclear spin relaxation becomes a mixture of the second-order nuclear quadrupole transitions and direct barrier-less transitions to other $s$-wave states.

When both molecules are in the state with $M_\text{Rb} = -\tfrac{3}{2} $ and $M_\text{Cs} = -\tfrac{5}{2}$, there are multiple $s$-wave channels open at all collision energies. The nuclear spin transitions do not have to change $L$ and the cross section for nuclear spin relaxation is an order of magnitude larger than for the maximally stretched state. It can also be seen that the nuclear quadrupole coupling is no longer dominant, as at low collision energy the is no significant change in the inelastic cross section when the quadrupole moments are set zero.

%
% MECHANISM
%
% ----
% figure 3, spin-relaxation cross sections with various hyperfine parameters turned off
%
%
% ----

\begin{figure}
\includegraphics[width=0.45\textwidth]{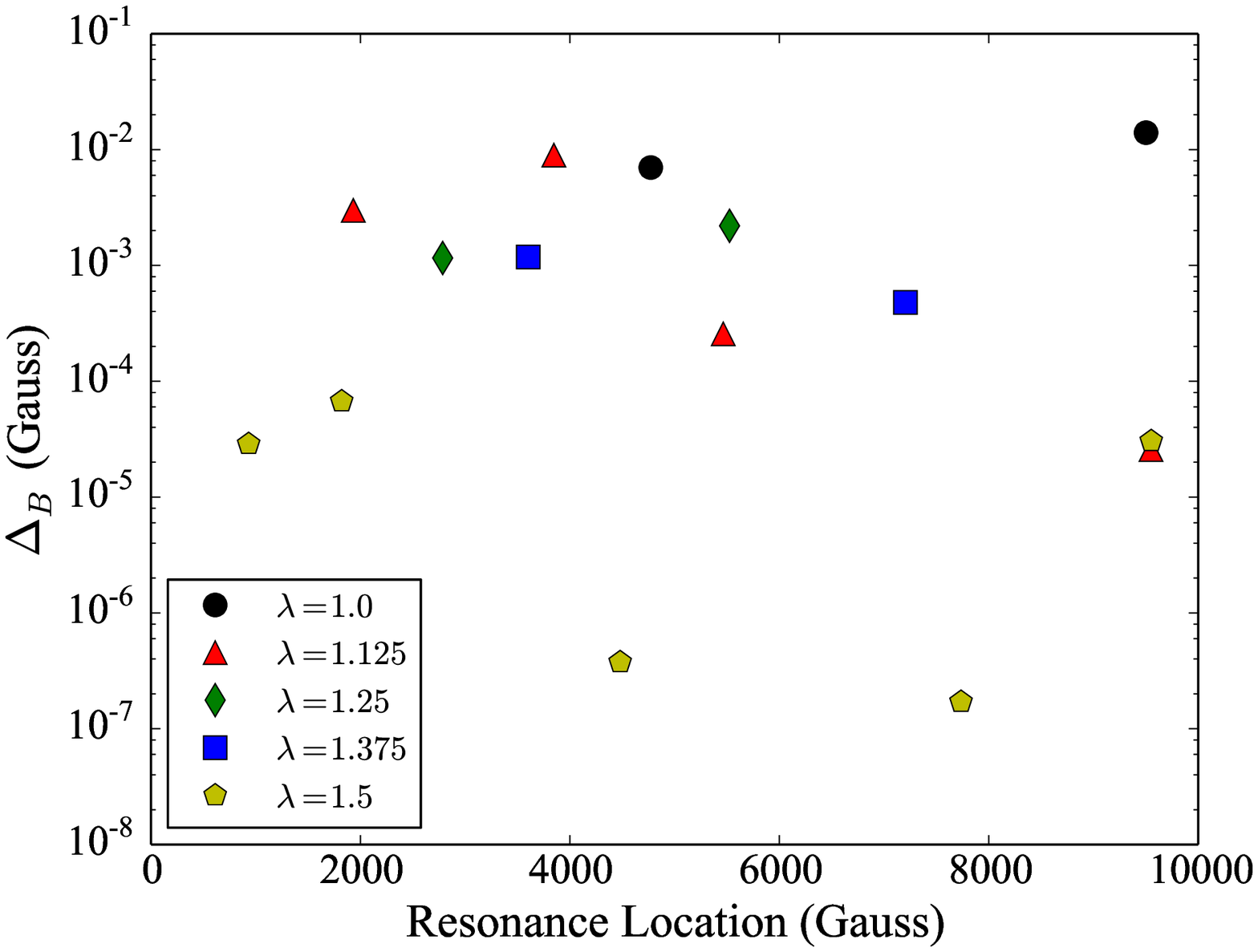}
\includegraphics[width=0.45\textwidth]{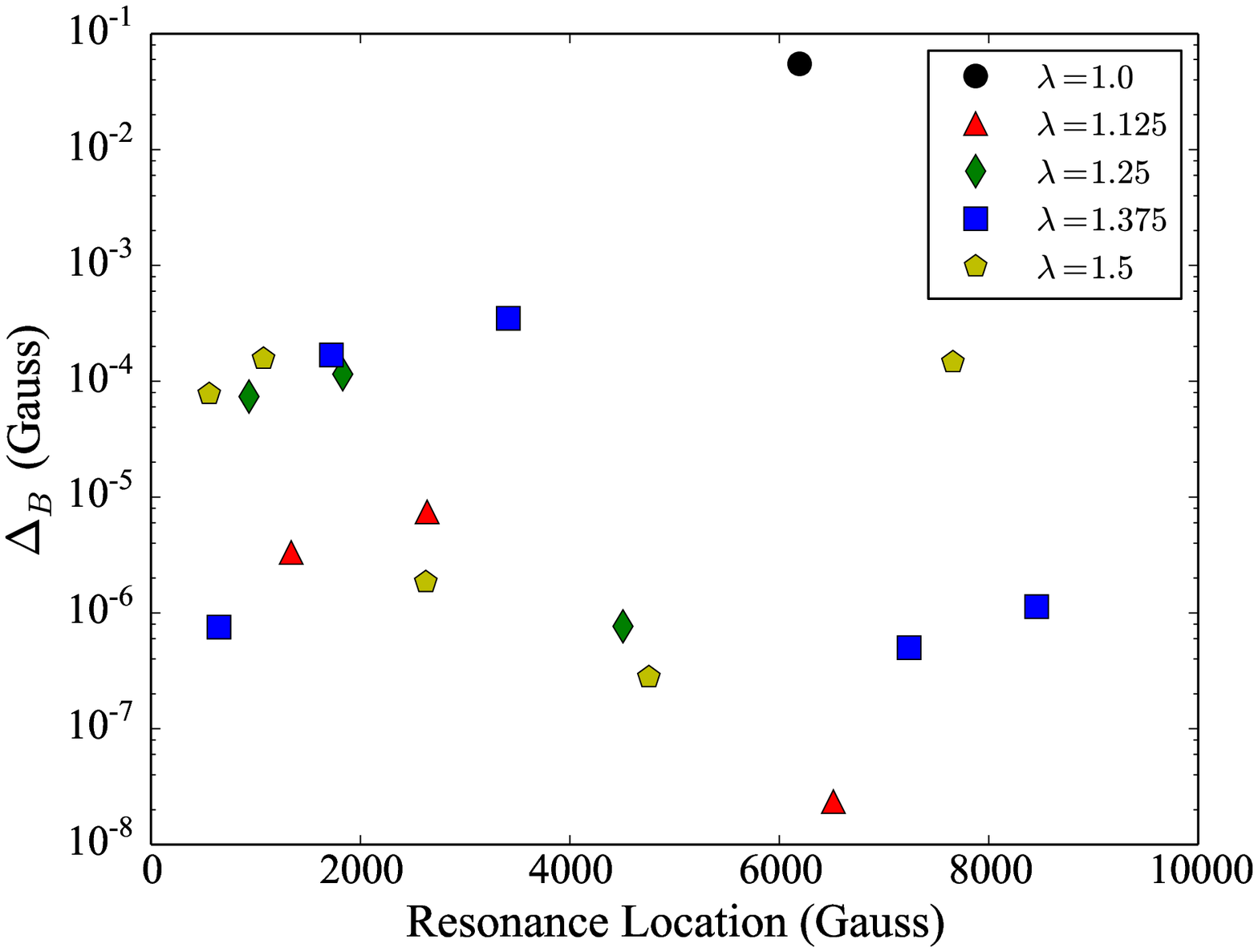}
\caption{ 
(color online) The locations and widths $\Delta_B$ of magnetic Feshbach resonances for collisions of RbCs molecules in the state $(M_{\rm Rb}=\tfrac{3}{2},M_\text{Cs}=\tfrac{7}{2})$ computed with various  interaction potential strengths $\lambda$, for the basis sets $N\le 2$, $L\le 4$ (upper panel) and $N\le3$, $L\le2$ (lower panel). 
\label{fig:resonances}
}
\end{figure}

The couplings responsible for nuclear spin relaxation are also responsible for magnetic Feshbach resonances. To illustrate this, we calculate the magnetic Feshbach resonances for molecules prepared in the lowest-energy spin-stretched state $(M_{\rm Rb}=\tfrac{3}{2},M_\text{Cs}=\tfrac{7}{2}$, see Figure 1). The calculations are performed for a collision energy of 1 nK and the magnetic field range between 0 and 10000 Gauss.  The resonance widths are obtained by fitting the scattering length across the Feshbach resonance to the formula \cite{Moerdijk:1995}
\begin{equation}
a(B) = a_\text{bg}\left[ 1 - \frac{\Delta}{B-B_\text{res}}\right].
\end{equation}

The quantitative accuracy of the results presented is limited by the interaction potential surface and the basis set restrictions. While it is unlikely that increasing the basis set may decrease the resonance widths observed, it is necessary to confirm this. 
To examine the effect of the interaction potential variation on the width of the resonances, we scale the NH-NH potential by a linear factor, $V^\text{scaled} = \lambda \times V^\text{NH-NH}$. 
We computed the resonance locations and widths using several potential surfaces with different values of $\lambda$ and three basis sets:
$N\le2$, $L\le2$, $N\le2$, $L\le4$ and $N\le 3$, $L\le 2$. 
The calculation with the most restricted basis set $N\le2$, $L\le2$ yields the resonance widths on the order of  0.1 $\mu$G. Increasing the basis set increases the width of the observed resonances. 
Figure~\ref{fig:resonances} plots the resonance locations and widths computed with two basis sets:
 $N\le2$, $L\le4$ and $N\le 3$, $L\le 2$. Different symbols correspond to different potential surfaces. 
It can be seen from Figure~\ref{fig:resonances} that, although the location of the resonances is very sensitive to the potential energy surfaces, the widths of the widest resonances are not strongly dependent on $\lambda$. This indicates that the resonance widths are limited by the relatively weak nuclear-quadrupole interaction and not the anisotropy of the interaction potential. 
For the $N\le 3$, $L\le 2$ basis set the largest resonance widths are around 0.1 mG, with a single resonance that is 0.1 G wide.
For the $N\le2$, $L\le4$ basis set the largest resonance widths are around 1 to 10 mG wide. 
%This suggests that the basis set with $N\le2$ gives a reasonable description of the resonances induced by the quadrupole interactions, whereas the potential anisotropy is poorly represented by $L\le2$.

\section{Conclusions}

The main question we have aimed to answer in this study is, can the widths of magnetic Feshbach resonances in collisions of $^1\Sigma$ molecules -- such as RbCs -- be large enough to be of experimental utility? Magnetic Feshbach resonances are routinely used to control interactions between ultracold alkali metal atoms. Current technologies permit the experimental resolution of magnetic Feshbach resonances with widths on the order of $1$ mGauss.  Our calculations show that a combination of intra-molecular hyperfine interactions with inter-molecular electrostatic interactions can lead to Feshbach resonances with widths up to 10 mGauss.

We note that the scattering of ultracold alkali metal dimers is expected to be strongly influenced by multiple resonances arising not only from the hyperfine structure of molecules but also from the dense spectrum of ro-vibrational states \cite{Mayle:2012,Mayle:2013}. We do not observe the large number of resonances predicted in Ref. \cite{Mayle:2013} because our calculations use a very limited basis set of rotational states and no vibrationally excited states of molecules. If the density of resonances in molecule - molecule collisions is large enough, the scattering of ultracold molecules may become predominantly resonant, leading to the formation of long-lived collision complexes and possible removal of molecules from ultracold gas through three-body recombination. These losses may preclude the magnetic field control of ultracold collisions of molecules. The likelihood of these detrimental processes is still a subject of debate.
It will depend on the density of the ro-vibrational resonances, which should generally be smaller than the density of the corresponding ro-vibrational states due to non-ergodic effects.

%Interestingly, the number of resonances we observed is rather scant. This is in opposition to the main assumption of the recent theory by Mayle and coworkers \cite{Mayle:2013} suggesting that strong interactions between molecules must lead to the appearance of a large number of resonances. In order to prove or disprove the assumption of Ref. \cite{Mayle:2013}, it is necessary to perform calculations with more extended basis sets, including more ro-vibrational states. However, it appears to be unlikely that the number of resonances will be anywhere close to the number of molecular states as was assumed in Ref. \cite{Mayle:2013}. Our present calculations show that the scattering dynamics involving about 4000 channels leads to the appearance of only about five resonances in the magnetic field interval between 0 and 10000 Gauss. 

%Finally, our calculations demonstrate that the cross sections for nuclear spin relaxation in molecule - molecule collisions can be significant. We obtain the cross sections for nuclear spin relaxation that are about six orders of magnitude smaller than the cross sections for elastic scattering.  Due to the restricted basis set and the model interaction potential used in the calculations, this should be considered as the lower limit. It is reasonable to expect that increasing the strength of the interaction potential and the number of basis states must increase the strength and number of couplings between the hyperfine states, leading to larger probabilities of nuclear spin relaxation. 

  The results presented here should be useful for the non-ergodic corrections in the statistical estimates of the density of resonances in collisions of molecules with hyperfine structure \cite{Mayle:2013}.  For example, our results show that the manifold of hyperfine states displayed in Figure 1 participating in the scattering dynamics encompassing about 4000 collision channels leads to the appearance of only about five resonances in the magnetic field interval between 0 and 10000 Gauss.  The next step in the analysis of resonant scattering of molecules should be a quantum calculation including a large number of rotational and vibrational states in addition to the hyperfine structure of molecules. We hope that the present work will stimulate further theoretical work of hyperfine structure effects in collisions of $^1\Sigma$ molecules as well as the experimental studies of ultracold collisions of $^1\Sigma$ molecules in a magnetic field.

%
% acknowlegdements
%

\section{Acknowledgments}

This work was supported by NSERC of Canada

%
%\bibliography{alisdair-references}

%\bibliography{/Users/alisdair/Dropbox/bibtex/attostuff,/Users/alisdair/Dropbox/bibtex/thref,/Users/alisdair/Dropbox/bibtex/all,/Users/alisdair/Dropbox/bibtex/molpro,/Users/alisdair/Dropbox/bibtex/AOGWbib}

\end{document}